\documentclass[preprint2]{aastex}

\usepackage{color}
\usepackage{ulem}

\def\Ms{$M_{\odot}$}

\shorttitle{FAST Discovered 24 Globular Cluster Pulsars}
\shortauthors{Pan et al.}

\begin{document}

\title{FAST Globular Cluster Pulsar Survey: Twenty-Four Pulsars Discovered in Fifteen Globular Clusters}

\author{Zhichen Pan\altaffilmark{1, 2, 3, *}, Lei Qian\altaffilmark{1, 2, 3}, Xiaoyun Ma\altaffilmark{1, 2}, Kuo Liu\altaffilmark{4}, Lin Wang\altaffilmark{1, 2, 3}, Jintao Luo\altaffilmark{5}, Zhen Yan\altaffilmark{6}, Scott Ransom\altaffilmark{7}, Duncan Lorimer\altaffilmark{8, 9}, Di Li\altaffilmark{1, 2, 3}, Peng Jiang\altaffilmark{1, 2, 3, *}}

\affil{$^1$National Astronomical Observatories, Chinese Academy of Sciences, 20A Datun Road, Chaoyang District, Beijing, 100101, China}
\email{panzc@bao.ac.cn, pjiang@bao.ac.cn}
\affil{$^2$CAS Key Laboratory of FAST, National Astronomical Observatories, Chinese Academy of Sciences, Beijing 100101, China}
\affil{$^3$College of Astronomy and Space Sciences, University of Chinese Academy of Sciences, Beijing 100049, China}
\affil{$^4$Max-Plank-Institut f{\"u}r Radioastronomie, Auf dem H{\"u}gel 69, Bonn, D-53121, Germany}
\affil{$^5$National Time Service Center, Chinese Academy of Sciences, Xi'an 710600, China}
\affil{$^6$Shanghai Astronomical Observatories, Chinese Academy of Sciences, Shanghai, 20030}
\affil{$^7$National Radio Astronomy Observatory, Charlottesville, VA 22903, USA}
\affil{$^8$National Time Service Center, Chinese Academy of Sciences, Xi'an 710600, China}
\affil{$^9$Center for Gravitational Waves and Cosmology, West Virginia University, Chestnut Ridge Research Building, Morgantown, WV 26505}

\begin{abstract}
  We present the discovery of 24 pulsars in 15 Globular Clusters (GCs) using the Five-hundred-meter Aperture Spherical radio Telescope (FAST).
  These include the first pulsar discoveries in M2, M10, and M14.
  Most of the new systems are either confirmed or likely members of binary systems.
  M53C, NGC6517H and I are the only three pulsars confirmed to be isolated.
  M14A is a black widow pulsar with an orbital period of 5.5 hours and a minimum companion mass of 0.016 \Ms.
  M14E is an eclipsing binary pulsar with an orbital period of 20.3 hours.
  With the other 8 discoveries that have been reported elsewhere,
  in total 32 GC pulsars have been discovered by FAST so far.
  In addition, We detected M3A twice.
  This was enough to determine that it is a black widow pulsar with an orbital period of 3.3 hours and a minimum companion mass of 0.0125 \Ms.
\end{abstract}


\section{Introduction}

Globular Clusters (GCs) harbour a large population of millisecond pulsars (MSPs).
Some of these can be quite exotic, e.g., the fastest spinning pulsar J1748-2446ad (Hessels et al. 2006).
This system is part of a large population of eclipsing binary pulsars,
most with orbital periods less than one day (see e.g., Ridolfi et al. 2021 for some recent examples),
and even a triple system of a pulsar together with a white dwarf and a planet companion (Thorsett et al. 1999).
The exotic systems in globular clusters also include the millisecond pulsars with eccentric orbits,
in particular of the systems with very massive companions:
these clearly result from exchange encounters that happened after the pulsars were fully recycled in LMXBs (Low Mass X-ray Binaries),
while nothing like them exists in the Galactic disk.
Some examples are:
NGC 1851A (orbital eccentricity 0.89, companion mass 1.22$^{+0.06}_{-0.05}$ \Ms, Ridolfi et al. 2019),
NGC 6544B (orbital eccentricity 0.75, companion mass 1.2064(20) \Ms, Lynch et al. 2012),
NGC 6652A (orbital eccentricity 0.97, companion mass 0.74 \Ms, DeCesar et al. 2015)
NGC 6624G (orbital eccentricity 0.38, companion mass 0.53$^{+1.30}_{-0.09}$ \Ms, Ridolfi et al. 2021),
and M15C (orbital eccentricity 0.68, companion mass 1.345$\pm$0.010 \Ms, Jacoby et al. 2006).
The total 230 pulsars in 36 GCs\footnote{\url{http://www.naic.edu/~pfreire/GCpsr.html}, keep increasing} till May 2021 are the result of GC surveys made with the largest existing radio telescopes
such as Parkes (e.g., tens pulsars discovered in 47 Tucanae, Manchester et al. 1991; Robinson et al. 1995; Camilo et al. 2000; Pan et al. 2016; Rildolfi et al. 2021),
Arecibo (e.g., 11 new pulsars discovered in a survey of 22 GCs, Hessels et al. 2007),
the Green Bank Telescope (e.g., the jackpot Terzan 5, Ransom et al. 2005 and Prager et al. 2017),
MeerKAT\footnote{http://trapum.org/discoveries.html},
and Giant Metrewave Radio Telescope (GMRT, Freire et al. 2004).

Since the commissioning of Five-hundred-meter Aperture Spherical radio Telescope (FAST, Nan et al. 2011; Jiang et al. 2019),
GCs have been important targets for pulsar searches with FAST.
Pan et al. (2020) reported an eclipsing binary discovered in M92.
This pulsar was discovered in a 500-700~MHz subband of the single beam ultra wide band receiver covering 270-1620~MHz with an 0.5-hr observation.
With the same receiver, Wang et al. (2020) reported the FAST discovery of M13F,
which might be an extremely faint and scintillating pulsar,
lately shown to be a high-mass neutron star (Cadelano et al. 2020); they also confirmed that M13E is a back widow system.
More GC pulsar discoveries were done with the 19-beam L-band receiver which replaced the single beam ultra wide band receiver on May of 2018.
Pan et al. (2021, accepted by RAA) discovered three faint isolated millisecond pulsars in NGC 6517.
The detection of short-orbit binaries with acceleration search (Ransom, Eikenberry \& Middleditch 2002) benefits from FAST's high sensitivity, too.
For example, the NGC 6712A (Yan et al. submitted to ApJ),
a black widow with an orbital period of 0.15 days,
was discovered in a 4-min segment from the FAST data with the acceleration search.
We believe that it is bright enough to be detected by GBT, but may have been missed due to its short orbit.

Until the December of 2021, we have searched 15 GCs with 24 new pulsars discovered and updated timing solutions for several previously discovered pulsars.
In Section 2, we describe the observation and the pulsar searching method.
Timing results are presented in Section 3.
Section 4 and 5 are the discussion and conclusion, respectively.

\section{Observation and Data Reduction} \label{sec:obs}

As a part of the SP$^2$ preject\footnote{Search of Pulsars in Special Population, \url{https://crafts.bao.ac.cn/pulsar/SP2/}},
the FAST GC pulsar survey\footnote{\url{https://fast.bao.ac.cn/cms/article/65/}} was started in 2018.
The FAST 19-beam receiver, which covers a frequency range of 1.0 to 1.5~GHz, was used for these observations.
Because GC pulsars are close to the centers normally, the data from the center beam were only recorded.
The data were channelized into 4096 channels, corresponding to 0.122~MHz channel width.
The signals were 8-bit sampled from two polarizations which are from two orthogonal dipoles. 
The sampling time is 49.152~$\mu$s. 
The system temperature is $\rm \sim$24~K (Jiang et al. 2020), and the beam size is $\rm \sim$2.9' at 1.4~GHz.
There are 45 GCs in FAST sky.
All 11 GCs with previously known pulsars were observed first.
In order to maximize the search sensitivity, these GCs were observed with the longest allowed integration time. 
The other four GCs, M14, M10, M2, and NGC 6712, in which we also found new pulsars during FAST commissioning, were also included.
Table \ref{GCs} shows more details of these observations carried out.
The search sensitivities in Table \ref{GCs} were calculated using the radiometer equation as follows (e.g., Lynch et al. 2011)
  \begin{equation}
S_{\rm min} = \beta \displaystyle  \frac{(S/N_{\rm min}) T_{\rm sys}}{G \sqrt{n_{\rm p} t_{\rm int} \Delta f}}\sqrt{\frac{W}{P-W}},
    \label{eq_sensitivity}
  \end{equation}
in which $\beta$ is the sampling efficiency and equals to 1 for our 8-bit recording system.
$W$ is the width of the pulsar profile, and we used 10\% of the pulse.
The minimum signal-to-noise ratio ($S/N)_{min}$ in our search is 10.
The system temperature ($T_{sys}$) is 24~K.
The antenna gain, G, is 16~K~Jy$^{-1}$.
The number of polarizations ($n_{p}$) is 2.
The t$_{int}$ is the integration time in the unit of second.
The $\Delta f$ is the bandwidth in the unit of MHz, 
here is 300~MHz,
as in our radio-frequency interference (RFI) masking $\rm \sim$25\% channels were masked as RFIs and will not be used in further process.

\begin{table*}[htpb]
\centering
\caption{Observation and Data Reduction Details. The observation date and length are for the longest observation done with the corresponding GC.}
\label{GCs}
\begin{tabular}{ccccccc}
\hline
GC name  & RA              &  DEC         &  Observation Date  & Observation   &   DM Range from  &   Sensitivity \\
         &   (J2000)       &  (J2000)     &          & Length (hr)             &   Known Pulsars  &  $\mu$Jy       \\
\hline
M2  (NGC7089)    &  21:33:27    &  -00:49:23.5 &  2020-01-04          &    2.0              &   43.3-44.1  &  0.76  \\
M3  (NGC5272)    &  13:42:12    &  +28:22:38.2 &  2019-12-14          &    4.5              &   26.1-26.5  &  0.51  \\
M5  (NGC5904)    &  15:18:33    &  +02:04:51.7 &  2019-12-11          &    4.0              &   29.3-30.1  &  0.54  \\
M10  (NGC6254)   &  16:57:09    &  -06:04:01.1 &  2020-01-05          &    3.0              &   43.4-43.9  &  0.62  \\
M13 (NGC6205)    &  16:41:41    &  +36:27:35.5 &  2020-12-27          &    5.0              &   30.1-30.5  &  0.16  \\
M14 (NGC6402)    &  17:37:36    &  -03:14:45.3 &  2019-07-17          &    2.0              &   78.8-82.1  &  0.76  \\
M15 (NGC7078)    &  21:29:58    &  +12:10:01.2 &  2020-12-21          &    5.0              &   65.5-67.7  &  0.48  \\
M53 (NGC5024)    &  13:12:55    &  +18:10:05.4 &  2019-11-30          &    5.0              &   24.0-26.1  &  0.48    \\
M71 (NGC6838)    &  19:53:46    &  +18:46:45.1 &  2019-12-12          &    5.0              &   117.4      &  0.48  \\
M92 (NGC6341)    &  17:17:07    &  +43:08:09.4 &  2020-03-12          &    3.6              &   35.45      &  0.57  \\
NGC6517          &  18:01:51    &  -08:57:31.6 &  2020-01-23          &    2.5              &   174.7-185.3&  0.68   \\
NGC6539          &  18:04:50    &  -07:35:09.1 &  2021-02-10          &    2.2              &   186.38     &  0.72   \\
NGC6712          &  18:53:04    &  -08:42:22.0 &  2019-07-21          &    2.0              &   155.2      &  0.76   \\
NGC6749          &  19:05:15    &  +01:54:03.0 &  2019-12-10          &    3.0              &   192.0-193.7&  0.62   \\
NGC6760          &  19:11:12    &  +01:01:49.7 &  2019-12-06          &    4.0              &   196.7-202.7&  0.54  \\
\hline
\end{tabular}
\end{table*}

The RPPPS package (Yu et al. 2020),
which is a parallel pipeline base on PRESTO (Ransom 2001, Ransom 2002, Ransom 2003), was used in the search.
The data were dedispersed with DM ranges (see in Table \ref{GCs}) a few times wider than those from known pulsars in the GC.
For GCs without previously known pulsars,
the DM ranges were initially set to be twice of the upper limits of the YMW16 model (Yao, Manchester \& Wang 2017) predictions on the direction of GCs.
The acceleration search with a summation of up to 32 Fourier harmonics was used for periodic signal search.
In order to detect both faint isolated pulsars and binaires, the zmax values were set to be 20 and 1200, respectively.
The observations normally lasted for more than 2 hours.
Thus, even with a zmax value of 1200,
a very limited range of acceleration was searched\footnote{The average acceleration of the pulsar in the integration can be calculated as $a = z \rm{c}/(T^2 f_0)$ (Ransom et al. 2001), in which $c$ is the speed of light, $T$ is the total integration time, $f_0$ is the pulsar spin frequency. If the total integration time was 5 hour, using 20/1200 as the zmax value corresponded to an acceleration of 0.09/5.5 m s$^{-2}$ for a 200 Hz signal or 0.04/2.2 m s$^{-2}$ for a 500 Hz signal.},
and thus the search sensitivity could be significantly reduced for pulsars with orbital period less than one day.
Now, we are processing segments for the GC data to re-detect these pulsars and search for new pulsars.
We will report results elsewhere.
To examining the candidates obtained or rejecting those unlikely to be pulsars and to combine multiple detections of the same signal,
both the \texttt{ACCEL\underline{ }sift.py} routine from \textsc{PRESTO} and our sifting code (\texttt{Jinglepulsar}\footnote{\url{https://www.github.com/jinglepulsar}}, see Pan et al. accepted by RAA) were used.
For candidates from both \texttt{ACCEL\underline{ }sift.py} and \texttt{Jinglepulsar},
we folded the dedispersed time series, filtered out RFI detections, and removed known pulsar harmonics.
The DM values, periods and accelerations of the remaining candidates were then used to fold the search data where they were discovered;
the resulting plots were visually inspected for pulsar detections.
As results, 24 new GC pulsars and all the previously FAST discovered pulsars (M92A, M13F, NGC5627E to G, and NGC6712A) were re-detected. 
Figure~\ref{psr_plots1} and \ref{psr_plots2} show the pulse profiles of the 24 new GC pulsars and Table \ref{discoveries} presents a summary of all the 30 GC pulsars discovered by FAST.
We will discuss these discoveries in Section 4.
Note that the most recent discoveries of M5G and M12A will be reported in separate papers and thus not listed here.

\begin{figure*}
\centering
    \includegraphics[width=10cm]{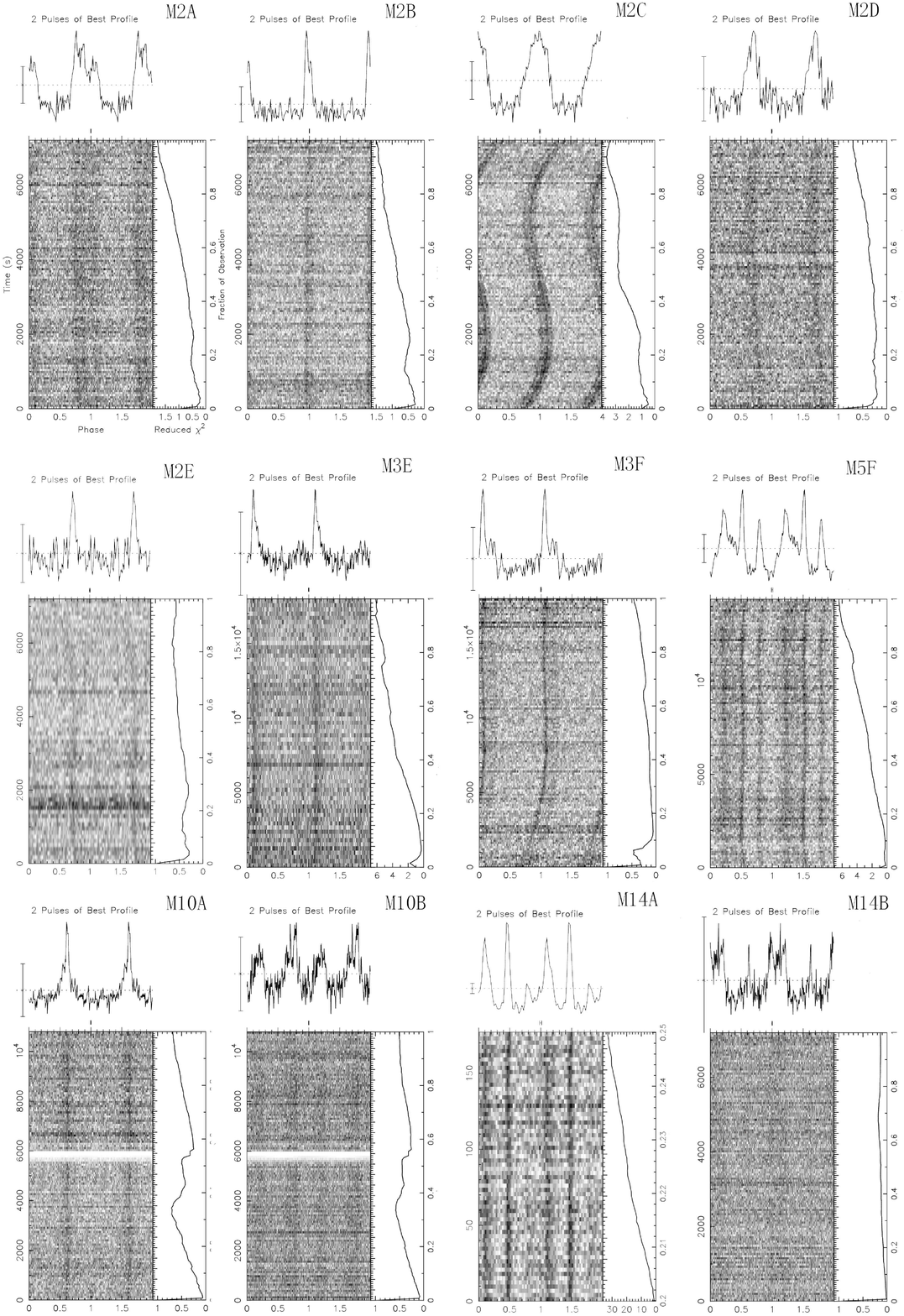}
    \caption{The 24 GC pulsars discovered using FAST. Pulsars from left to right and upper to bottom are presented in the same order as in Table Table~\ref{discoveries}. M92A, M13F, NGC6517E to G, and NGC6712A are previously discovered and not included in these figures.}
    \label{psr_plots1}
\end{figure*}

\begin{figure*}
\centering
    \includegraphics[width=10cm]{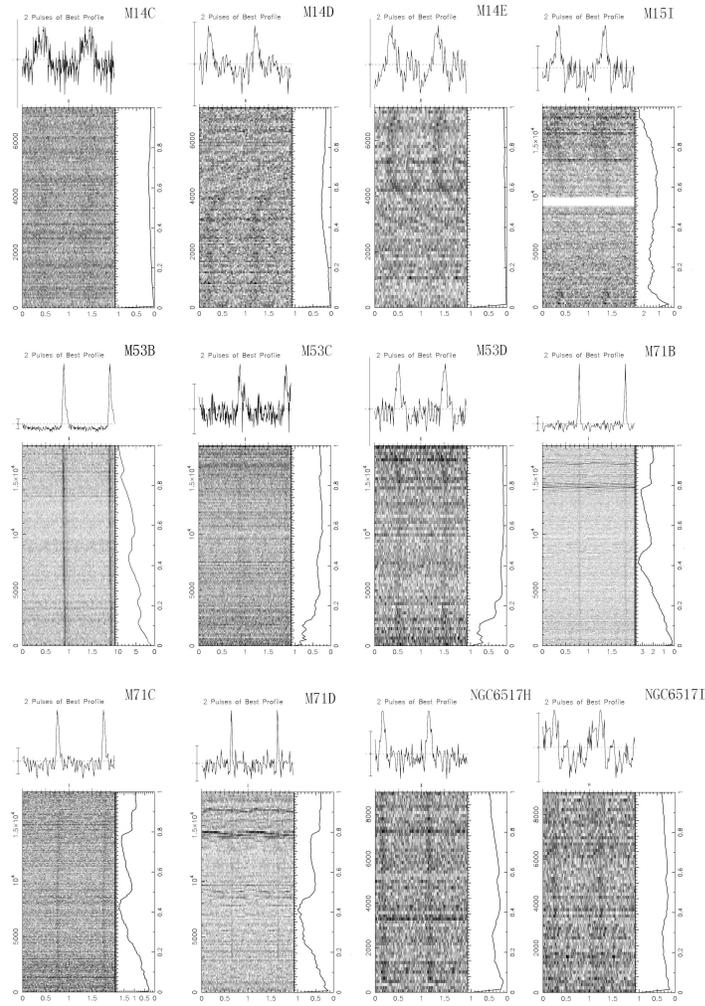}
    \caption{Continuation of Figure~\ref{psr_plots1}.}
    \label{psr_plots2}
\end{figure*}

\begin{table*}[htpb]
\centering
\caption{The 30 GC pulsars discovered using FAST.
The DM values come from DM searching and optimizing results during pulsar search.
The pulsar M10A and M71D (with question marks) are very good candidates but need more observations to confirm.}
\label{discoveries}
\begin{tabular}{ccccc}
\hline
Name           &   P0 (ms)    &   DM (pc cm$^{-3}$)  &  Notes              &  References  \\
\hline
M2A            &   10.15      &   43.3               &  Binary             &  \\
M2B            &   6.97       &   43.8               &  Binary             &  \\
M2C            &   3.00       &   44.1               &  Binary             &  \\
M2D            &   4.22       &   43.6               &  Binary             &  \\
M2E            &   3.70       &   43.8               &  Binary             &  \\
M3E            &   5.47       &   26.541             &  Binary             &  \\
M3F            &   4.40       &   26.467             &  Binary             &  \\
M5F            &   2.65       &   29.4               &  Binary             &  \\
M10A$^{?}$     &   4.73       &   43.9               &  Need Confirmation  &  \\
M10B           &   7.35       &   43.355             &  Possible Binary    &  \\
M13F           &   3.00       &   30.366             &  Binary             &  Wang et al. (2020) \\
M14A           &   1.98       &   82.10              &  Black Widow        &  \\
M14B           &   8.52       &   81.0               &  Binary             &  \\
M14C           &   8.46       &   80.0               &  Binary             &  \\
M14D           &   2.89       &   78.8               &  Redback            &  \\
M14E           &   2.28       &   80.4               &  Eclipsing Redback  &  \\
M15I           &   5.12       &   67.3               &  Possible Binary    &  \\
M53B           &   6.07       &   25.959             &  Binary             &  \\
M53C           &   12.53      &   26.106             &  Isolated           &  \\
M53D           &   6.07       &   24.6               &  Binary             &  \\
M71B           &   79.90      &   119.0              &  Possible Binary    &  \\
M71C           &   28.93      &   116.2              &  Possible Binary    &  \\
M71D$^{?}$     &   100.67     &   119.038            &  Need Confirmation  &  \\
M92A           &   3.16       &   35.45              &  Eclipsing Redback  &  Pan et al. (2020) \\
NGC6517E       &   7.60       &   183.29             &  Isolated           &  Pan et al. accetped by RAA\\
NGC6517F       &   24.89      &   183.71             &  Isolated           &  Pan et al. accetped by RAA\\
MGC6517G       &   51.59      &   185.3              &  Isolated           &  Pan et al. accetped by RAA\\
NGC6517H       &   5.64       &   179.6              &  Isolated           &  \\
NGC6517I       &   3.25       &   177.8              &  Isolated           &  \\
NGC6712A       &   2.15       &   155.2              &  Black Widow        &  Yan et al. submitted to ApJ\\
\hline
\end{tabular}
\end{table*}

\section{Results and Discussions}

In total, 32 GC pulsars have been discovered with FAST up to now,
which doubles the number of known GC pulsars in the FAST sky (from 31 to 63).
Below we discuss the 24 new discoveries made in this work.
Because the M5G was discovered by others (even it was also re-detected by us) and M12A was just discovered,
the discussion on M5G and M12A will be presented in other papers.

\subsection{Timing Results}

After the new discoveries were made, we searched the FAST archival data to confirm and time these pulsars.
With enough archived data, the phase connected timing solutions of M14A and NGC65117H were obtained (see Table \ref{timing_parameters}). 

\begin{table*}
\caption{Timing solutions of two newly discovered pulsars.}
\label{timing_parameters}
\begin{tabular}{ccc}
\hline\hline
Pulsar name                &      M14A     &     NGC6517H   \\
\hline
MJD range        \dotfill             & 58458---59154  & 58659---59105  \\
Data span (days)   \dotfill           &      697       &      446       \\
Number of TOAs      \dotfill          &       19       &       35       \\
Timing residuals r.m.s. ($\mu$s). \dotfill &      0.67      &     49.95       \\
\hline
\multicolumn{3}{c}{Measured quantities} \\
\hline
Right ascension, RA (J2000)  \dotfill     &  17:37:35.88794(3) &  18:01:52.5895(9)\\
Declination, DEC (J2000)      \dotfill    &  -03:14:34.775(1)  &  -08:57:53.15(5)\\
Spin frequency (Hz)     \dotfill                   &  505.05656417562(4) &  177.2193886845(2)\\
Spin frequency derivative (s$^{-2}$)   \dotfill    &  -2.4376(3)e-14     &  -1.210(7)e-14\\
Orbital Period, P$_b$ (days)          \dotfill            &   0.2278292041(9)   &  ---\\
Epoch of Periastron passage, T$_0$ (MJD)   \dotfill     &   58681.7064745(9)  &  ---\\
Projected Semi-major Axis, $\chi_p$ (lt-s)  \dotfill &   0.047110(1)       &  ---\\
\hline
\multicolumn{3}{c}{Set quantities} \\
\hline
Reference epoch (MJD)        \dotfill             &  58900             &  58868\\
DM (cm$^{-3}$~pc)           \dotfill              &  82.1              &  179.6\\
Orbital eccentricity, $e$    \dotfill             &  0                 &  ---\\
Longitude of periastron, $\omega$ (deg) \dotfill  &  0                 &  ---\\
Solar System ephemeris         \dotfill           &  DE200             &  DE200\\
Binary model          \dotfill                    &  BT                &  ---\\
\hline
\end{tabular}
\end{table*}

M3A were detected twice in observations with length of 4.5 and 5 hours, respectively.
While its orbital period is shorter than the length of any of the two observations,
we successfully determined its previously unknown orbital parameters with the data obtained on 14$^{th}$, December 2019, 
showing that it is a black widow system with an orbital period of 0.1359 days (3.26 hours) and a minimum companion mass of 0.0125 \Ms.
Because the gap between the only two M3A observations is almost one year,
we cannot determine a phase-coherent timing solution for this system.
The timing solutions of M92A and NGC6517B were also updated.
As a redback system, the orbital frequency and its derivative of M92A were measured.
The NGC6517B solution is consistent with previous ones from Lynch et al. (2011).
Table \ref{timing_known} are the timing solutions of these three pulsars.

\begin{table*}
\centering
\caption{Updated Timing solutions for previous discovered pulsar M3A, M92A and NGC6712B.
For M3A, RA, DEC, and pulse frequency derivative are set quantities, labeled by $^s$;
For NGC6517B, eccentricity and periastron are measured, labeled by $^m$.}
\label{timing_known}
\begin{footnotesize}
\begin{tabular}{cccc}
\hline\hline
\hline
Pulsar name                   &      M3A       &      M92A       &      NGC6517B       \\
\hline
MJD range      \dotfill              & 58830---58831  & 58351---58991   & 58659---59105  \\
Data span (days)     \dotfill        &      $<$1      &      641        &    447         \\
Number of TOAs     \dotfill          &     29         &     119         &     20       \\
Timing residual r.m.s. ($\mu$s) \dotfill &     20.82      &     8.01        &     3.74       \\
\hline
\multicolumn{4}{c}{Measured quantities} \\
\hline
Right Ascension, RA (J2000)  \dotfill     &  13:42:00.67$^s$    &  17:17:49640(4)      &  18:01:50.5642(4)   \\
Declination, DEC (J2000)     \dotfill     &  +28:22:31.4$^s$    &  +43:08:03.4806(6)   &  -08:57:32.87(2)\\
Pulse Frequency (Hz)     \dotfill                   & 392.967704(6)       &  316.48368670253(5)  &  34.5284928861(1)  \\
Pulse Frequency Derivative (s$^{-2}$)  \dotfill     & 0$^s$               &  -6.123(3)e-15       &  -2.628(7)e-15  \\
Orbital Period, P$_b$ (day)        \dotfill               & 0.13590(2)          &  ---                 &  59.836454(1)  \\
Orbital Frequency, F$_b$ (Hz)    \dotfill               &  ---                &  5.762033487(5)e-5   &  ---  \\
Orbital Frequency Derivative, $\dot{F}_b$ (Hz$^{-2}$)    \dotfill   &  ---                &  -5.7(2)e-20         &  ---  \\
Epoch of passage at Periastron, T$_0$ (MJD) \dotfill & 58831.084513(5)     &  58353.54908181(4)   &  54757.72304(9)  \\
Projected Semi-major Axis, $\chi_p$ (lt-s) \dotfill & 0.025609(6)         &  0.3987068(7)        &  33.87543(1)  \\
\hline
\multicolumn{4}{c}{Set quantities} \\
\hline
Reference epoch (MJD)       \dotfill              & 58685.702931      &   58390   & 58670.000000  \\
Dispersion measure, DM (cm$^{-3}$~pc)  \dotfill   & 26.5              &   35.45   &  182.39  \\
Orbital eccentricity, $e$        \dotfill         & 0                 &   0       &  0.0382258(7)$^m$ \\
Longitude of periastron, $\omega$ (deg) \dotfill  & 0                 &   0       &  -57.8914(6)$^m$  \\
\hline
\multicolumn{4}{c}{Timing model assumptions} \\
\hline
Solar System ephemeris model    \dotfill      & DE200         &  DE436   &  DE200  \\
Binary model        \dotfill          & BT            &  BTX     &  BT  \\
\hline
\end{tabular}
\end{footnotesize}
\end{table*}

\subsection{M2}
In total, five new pulsars were discovered in M2.
These are the first pulsars found in this cluster.
The DM values of these pulsars falls in the range of 43.3-44.1~pc~cm$^{-3}$, close to the one predicted from YMW16 (34.59~pc~cm$^{-3}$).
The variation of their observed spin periods indicates that they are all binary pulsars with  orbital periods of several days.
Additional observations are required to determine their orbital parameters.
The situation of M2 that all discovered pulsars are in binary systems,
is very similar to that of M3, M5, M14 and M62 where no more than one isolated pulsar has been found in each cluster (Lynch et al. 2012).
The very high percentage of binary system in the discovered pulsars of these clusters can be explained by their relatively low encounter rate per binary (Verbunt \& Freire, 2014),
which leaves a high surviving probability of a binary system once formed.

\subsection{M3}
There were four pulsars discovered in M3 prior to our survey. 
Of these, M3A did not have a timing solution, or even a known orbit, and M3C has not been confirmed.
From our FAST observations of M3, we managed to re-detect three of the known pulsars except M3C.
M3E was discovered at our first attempt and detected for six times out of ten observations on M3,
which is more frequently than the other pulsars in M3.
Its spin period is approximately 5.47\,ms, close to but still clearly different from that of M3D.
M3F was discovered in a later observation, but only seen twice out of ten epochs.
Both pulsars are in binary systems.
The flux density of all known pulsars in this cluster shows significant variation,
which is likely to be a result of interstellar scintillation.

The detection and timing of M3A was mentioned in the first subsection of this section.

\subsection{M5}

With one pulsar discovered by us, the number of known pulsars in this cluster is six.
The new pulsar, M5F, was found in an on-going search accounting for the second derivative of the spin period (i.e., jerk, Anderson \& Ransom 2018) as an additional check on some of the epoch data to the standard scheme described in Section~\ref{sec:obs}.
Its spin period is approximately 2.65\,ms, and a DM of 29.5~pc~cm$^{-3}$ which is close to all the other known pulsars.
With all the eight epoch observations on M5,
we managed to obtain the orbital parameters for M5F.
Its orbital period is approximately 1.61\,days, and the minimum companion mass is 0.2 \Ms,
indicating that M5F is likely to have a white-dwarf or low mass star as the companion.

\subsection{M10}
Two new pulsars were discovered in M10 (one need further confirmation).
They are the first pulsars found in this cluster.
The DM values of the two new pulsars, M10A and B, are approximately 43.9 and 43.4~pc~cm$^{-3}$, respectively.
These values are inconsistent with the prediction by the YMW16 model which gives $\sim$107~pc~cm$^{-3}$ for the location of M10.
This may be a result of the lack of pulsars detected in this region of the sky.
M10A was detected with high significance in our first observation as shown in Figure~\ref{discoveries},
and seen to exhibit period variation probably as a result of acceleration in a binary system.
However, it was not detected in the second observation on M10.
M10B was detected on both epochs,
with significantly different signal-to-noise ratios which is likely a consequence of interstellar scintillation.
Thus, the non-detection of M10A in the second epoch could be attributed to either scintillation or the orbital phase of the pulsar when observed which may have an impact on the sensitivity of our search.

\subsection{M14}
We discovered five pulsars in M14.
These are the first pulsars found in this cluster.
Their DM values range from 78.8-82.1~pc~cm$^{-3}$, different from the prediction of 140~pc~cm$^{-3}$ by YMW16 model.
M14A has a spin period of approximately 1.98\,ms, being the secondly fastest spinning GC black widow after Terzan 5O (1.68 ms, Ransom et al. 2005).
M14B and C have relatively wide pulse profiles,
and their spin periods vary slightly in different epochs, indicating they are binaries with orbital periods being of the order of a few days.
We can't obtain the phase connected timing solutions for M14D and E, but obtained their orbital paramters.
M14D has an orbital period of 0.74\,days and may exhibit eclipsing phenomenon.
M14E has an orbital period of 0.85\,days and shows clear eclipsing at particular orbital phase.
Considering their circular orbits, low-mass companions, and the eclipsing phenomena,
both M14D and E are likely to be redback pulsars.

\subsection{M15}
Our discovery of M15I is the first newly discovered pulsar in M15 in the past more than 20 years (Anderson, PhD thesis 1993).
It was detected in a 5-hr observation, which was interrupted in the middle due to the mechanical issue,
and the pulsar signal was seen only before and after the interruption.
The DM value of M15I is approximately 67.3 pc cm$^{-3}$, consistent with those known pulsars  M15.
We saw significant period variation during the entire observation,
indicating that M15I is likely to be in a binary system.

In our observations, M15A, B, D, E, and F are always detectable even the observation time is half an hour.
M15C was only detected in four observations done on October and November of 2018, December of 2019, and August of 2020.
The reason of missing its detections may be the relatively short orbit and/or farther from the GC center than other pulsars.
M15G was only detected once, with very faint signal.
The signal is with a DM value of 67.9~pc~cm$^{-3}$ and a spin period of 37.660171 ms, which is consistent with previous results.
M15H was detected when the length of the observation is 3 hours or longer, indicating that it is too faint so that can not be detected in shorter observations.

\subsection{M53}

M53 is the most distant GC that has known pulsars currently.
We managed to re-detect M53A and discover three new pulsars, M53B, C, and D which are all significantly fainter than M53A.
The DM values of M53 B, C, and D are 26.0, 26.1, and 24.6~pc~cm$^{-3}$, respectively,
all slightly larger than but close to the previously discovered M53A.
Now, 4 pulsars with similar DM values are in the line of sight to M53.
As they all have similar DM values, they should be in the GC M53.
M53B and D are found to be binary pulsars, while C is likely to be isolated.
M53B and C have been detected in all of the five 5-hour observations for M53.
Nevertheless, M53D is relatively faint and has been seen in only $\rm \sim$50\% of the observations rate.
The gaps between the five observations are up to several months, making it difficult to determine the phase-connected timing solution to these pulsars.

\subsection{M71}

In the FAST sky, M71 is the nearest  GC that has known pulsars.
From our observations on M71, we re-detected M71A which is an eclipsing black widow pulsar.
In addition, we discovered three new pulsars, namely M71B, C, and D.
All of them have relatively long spin periods which are 79.9 ms, 28.9 ms, and 100.7 ms, respectively.
Judging from their barycentric periods in different observations, M71B and C should be binaries.
M71D is the faintest one among them.
Though we discovered it in a 5-hr observation,
it was not seen in our second M71 observation which though only lasted for 0.5\,hr.
Thus, a confirmation of M71D is still needed.

Recently, Han et al. (RAA in press) reported a pulsar discovered 2.5 arcminutes to the center of M71.
The DM value of this pulsars is 113~pc~cm$^{-3}$ (the DM range of four currently known pulsars is 116.2-119.0~pc~cm$^{-3}$), slightly smaller than all the currently known pulsars in M71.
Because this pulsar is far away from our observation center, no doubt that such a signal was not detected in our pulsar search.
It is possible that this pulsar is at the edge of M71 thus has smaller DM value, or is located in the foreground and not related to the cluster.
Since the Terzan 5B (Lyne et al. 1990, 205~pc~cm$^{-3}$) was proved to be a foreground pulsar due to the significantly different DM value (Ransom et al. 2005, 234-243~pc~cm$^{-3}$), 
this can be the second example in determining a pulsar either a member of a GC or just in the line of sight.

\subsection{NGC6517}
As mentioned above, in previous work three isolated pulsars were discovered with the Jinglepulsar pipeline in NGC6517 (Pan et al., accepted by RAA).
Afterwards, the integration time of following observations were increased from 30 minutes to 2 hours.
This results in the discovery of two more isolated pulsars with both the \textsc{presto} and \texttt{Jinglepulsar} pipeline.
NGC6517H is a relatively bright isolated pulsar.
It has been detected for several times and from those detections we obtained its phase-connected timing solution.
NGC6517I is most likely to be an isolated pulsar as well due to the fact that its barycentric spin period are highly consistent in different observations.
However, it is too faint that it was only detected for three times so its phase-connected timing solution is still left to be determined.

\subsection{Other GCs without Pulsar Discoveries}

We did not discovered any new pulsars in other GCs including M13, M92, NGC6539, NGC6712, NGC6749, and NGC6760,
while each of which we observed at least once with the longest possible exposure.
M13 has 6 pulsars and we successfully detected all of them with no additional new pulsar discovered.
M92 has been monitored roughly once per month mainly in order to time M92A (Pan et al. 2020),
and no additional new pulsar has been discovered from these observations.
NGC6539A is the only pulsar in this GC and was very clearly seen in our observation, with no additional discovery, neither.
There are two pulsars in NGC6749.
NGC6749A, which is a binary, was detected in a 3-hour observation.
NGC6749B, which is not confirmed, was not confirmed by us, neither. 
There is one black widow and one isolated pulsar in NGC6760.
Both of them were detected with no additional candidates obtained. 
A black widow pulsar has been discovered and timed in NGC6712 by FAST,
the details are reported elsewhere (Yan et al., submitted to ApJL).
However, from more than 10 observations, we did not find any new pulsars.
While we also propose to observe these GCs in the future,
discoveries on these GCs may come from searching in the larger sky area of these GCs,
or more sophisticated acceleration search.

\section{Conclusions}

  Our conclusions are as follows:

  1, We present 24 new pulsars discovered using FAST.
  Till the March of 2021, 32 GC pulsars has been discovered from FAST data.
  The number of GC pulsars in FAST sky was doubled (from 31 to 63).

  2, These discoveries were possible mainly because of the high sensitivity of FAST.
  GCs such as M3, M5, M53, and M71, where pulsars had been found before with Arecibo,
  were observed with FAST for 2 or more hours with several new pulsar discoveries.
  These discoveries indicate that FAST reaches a significant higher sensitivity than that of the earlier Arecibo surveys.

  3, The discoveries from either FAST or other telescopes such as GBT and MeerKAT may affect the GC pulsar number predictions.

  4, With our discoveries, all five known pulsars in M2, six in M3, and five in M14 are members of binary systems. NGC6517 is the only GC where we have only discovered isolated pulsars were discovered.

\acknowledgments
This work is supported by the Basic Science Center Project of the National Nature Science Foundation of China (NSFC) under Grant Nos. 11703047, 11773041, U1931128, U2031119. 
KL acknowledges the financial support by the European Research Council for the ERC Synergy Grant BlackHoleCam under contract no. 610058. 
This work made use of data from the Five-hundred-meter Aperture Spherical radio Telescope (FAST).
FAST is a Chinese national mega-science facility, built and operated by the National Astronomical Observatories, Chinese Academy of Sciences (NAOC).
We appreciate all the people from FAST group for their support and assistance during the observations.
We thank the anonymous referee for all the comments which which the language and science are improved.
ZP is supoorted by the CAS "Light of West China" Program.
LQ is supported by the Youth Innovation Promotion Association of CAS (id. 2018075).

\bibliography{psrrefs}

\end{document}